\documentclass[a4paper,10pt]{article}
\usepackage[T1]{fontenc}      
\usepackage[utf8]{inputenc}   
\usepackage[english]{babel}   
\usepackage{url}              
\usepackage{bm}
\usepackage{amsmath}
\usepackage{amsfonts}
\usepackage{graphicx}
\usepackage[]{appendix}
\usepackage{listings,lstautogobble}

\usepackage{tabularx}
\usepackage{pdflscape}
\usepackage{longtable}
\usepackage{bigints}
\usepackage{amsmath}
\usepackage{amssymb}
\usepackage{float}
\usepackage[footnotesize,bf]{caption2}
\usepackage{bm}
\usepackage{geometry}
\usepackage{etoolbox}
\usepackage{amsthm}
\usepackage{amsmath}
\usepackage{xcolor}
\usepackage{algorithm}
\usepackage{algpseudocode}
\usepackage{bm}

\usepackage{booktabs}
\usepackage{threeparttable}
\usepackage{float}
\usepackage{url}

\usepackage{geometry}
\usepackage{graphicx}
\graphicspath{{img/}}

\usepackage{scalerel}
\usepackage{xcolor}

\usepackage{multirow}

\usepackage{algpseudocode}

\definecolor{blucite}{RGB}{12,127,172}

\usepackage[authoryear]{natbib}
\usepackage[colorlinks, citecolor=blucite]{hyperref}

\title{\LARGE \textbf{Discrimination performance in illness-death models with interval-censored disease data}\\\vspace{0.4cm}}
\author{\large  Marta Spreafico$^{1,2,\dagger,*}$, Anja J. Rueten-Budde$^{1,\dagger}$,  Hein Putter$^{2,1}$ and Marta Fiocco$^{1,3,2}$\\
	\vspace{-2mm}\\
		\small \textit{$^{1}$Mathematical Institute, Leiden University, Leiden 2333 CC, NL}\\
	\small \textit{$^{2}$Department of Biomedical Data Sciences, Leiden University Medical Center, Leiden 2333 ZA, NL}\\
	\small \textit{$^{3}$Trial and Data Center, Princess Máxima Center for Pediatric Oncology, Utrecht  3584 CS, NL}\\
		\small $^{\dagger}$These authors contributed equally to this article.\\ \vspace{2mm}\\ 
	\normalsize *Corresponding author: \href{mailto:m.spreafico@math.leidenuniv.nl}{\normalsize \texttt{m.spreafico@math.leidenuniv.nl}}}

\begin{document}
\normalsize
\maketitle

\vspace{1mm}
\noindent \textbf{Disclaimer}:  This is the accepted version of the manuscript that has been published in 
\textit{\href{https://journals.sagepub.com/doi/10.1177/09622802251412855}{Statistical Methods in Medical Research}}. 
Please cite the published article available at DOI: \href{https://journals.sagepub.com/doi/10.1177/09622802251412855}{10.1177/09622802251412855}.
\vspace{2mm}

\begin{abstract}
	\small
	In clinical studies, the illness-death model is often used to describe disease progression. A subject starts disease-free, may develop the disease and then die, or die directly.
	In clinical practice, disease can only be diagnosed at pre-specified follow-up visits, so the exact time of disease onset is often unknown, resulting in interval-censored data.
	This study examines the impact of ignoring this interval-censored nature of disease data on the discrimination performance of illness-death models, focusing on the time-specific Area Under the receiver operating characteristic Curve (AUC) in both incident/dynamic and cumulative/dynamic definitions.
	A simulation study with data simulated from Weibull transition hazards and disease state censored at regular intervals is conducted.
	Estimates are derived using different methods: the Cox model with a time-dependent binary disease marker, which ignores interval-censoring, and the illness-death model for interval-censored data estimated with three implementations - the piecewise-constant model from the \texttt{msm} package, the Weibull and M-spline models from the \texttt{SmoothHazard} package.  These methods are also applied to a dataset of 2232 patients with high-grade soft tissue sarcoma, where the interval-censored disease state is the post-operative development of distant metastases. The results suggest that, in the presence of interval-censored disease times, it is important to account for interval-censoring not only when estimating the parameters of the model but also when evaluating the discrimination performance of the disease.
	\vspace{2mm}
\end{abstract}

\vspace{1mm}
\normalsize

\maketitle                   

\section{Introduction}\label{sec-intro}
\begin{figure}
	\centering
	\includegraphics[width=8cm]{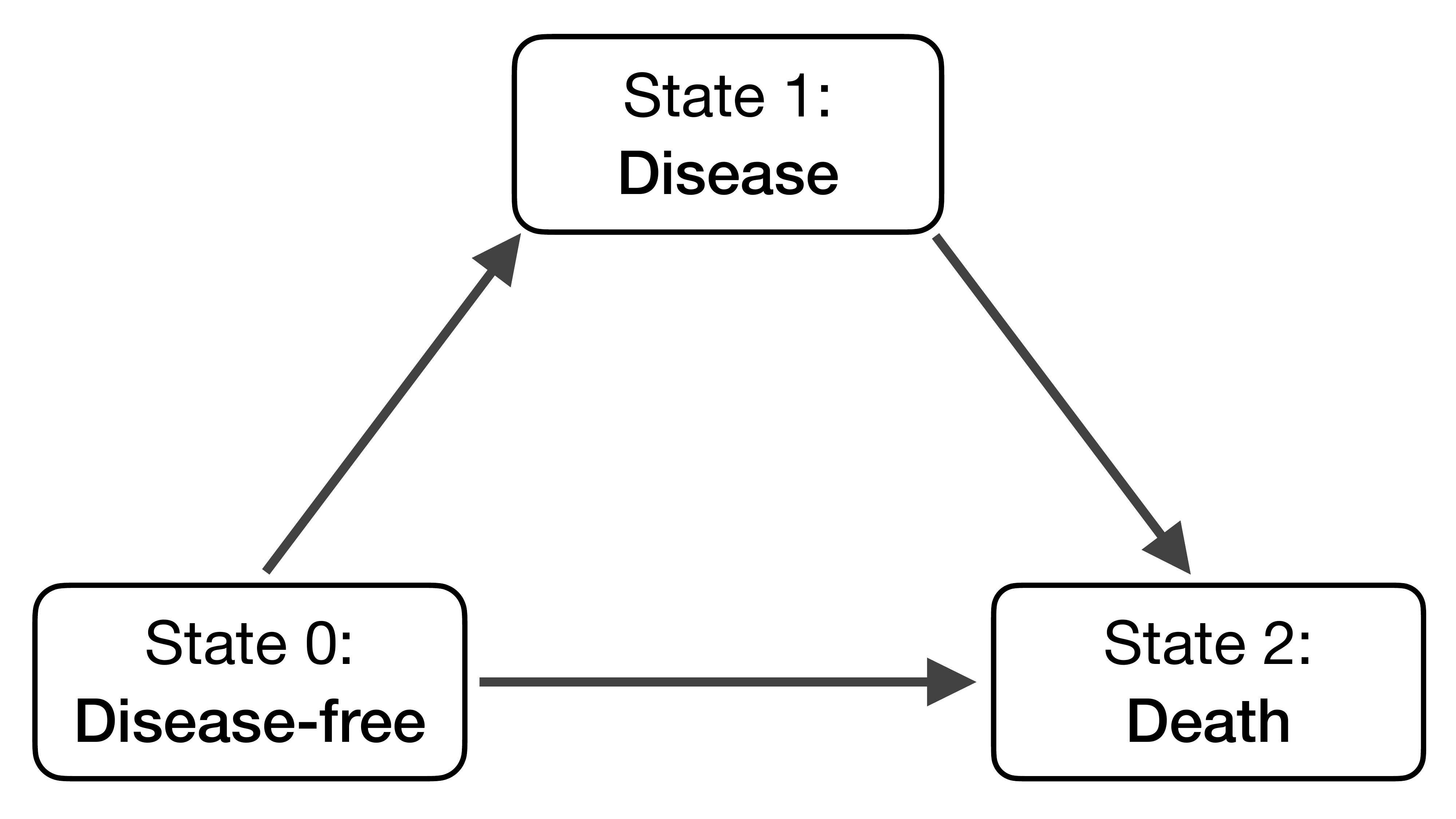}
	\caption{Illness-death model.}
	\label{idmodel}
\end{figure}

Survival analysis studies the distribution of time from an origin event to an event of interest \citep{survival-book}. It is often applied in the medical field where for example the time from diagnosis to death is studied. The intrinsic peculiarity of survival data is that they are generally incomplete: the event of interest cannot always be observed because it takes time to observe it. Data of individuals who did not experience the event of interest within a specific time window are hence right-censored. A frequently used method to study the effect of covariates on survival time is the Cox proportional hazards model \citep{cox}. In the medical field it is often applied to study the effect of risk factors on a single event such as death or disease progression. However, in practice disease progression may be described by more than one type of event. These more complex event structures can be modeled simultaneously using multi-state models \citep{tutorial,Cook2020,Andersen2023}. The most simple of such models is the illness-death model, which is described by three states (see Figure \ref{idmodel}): an individual is initially disease-free (state 0), he may then develop disease (state 1) and die (state 2) or he may die without disease. Like in the single event situation the Cox model can be used to model the effect of covariates on the transitions between states.

The illness-death model is applicable to a variety of disease settings; a problem arises, however, if the time of disease cannot be observed exactly. Often, disease can only be diagnosed at pre-specified follow-up times. An example lies in the care of patients with soft tissue sarcoma. After initial treatment by tumor removal surgery a patient may first develop distant metastases and then die. Metastases are diagnosed at pre-specified follow-up visits at which an X-ray of the patient is screened. If metastases are found, it is therefore only known that they appeared between the last negative screening and the first positive screening. This peculiar disease data contains two types of missing information: (1) the time of disease is only known to have happened between two visits, i.e., it is \textit{interval-censored}, (2) if the last disease screening prior to death or last recorded follow-up was negative the disease status of a patient between last screening and death or last recorded follow-up is unknown.

The illness-death model for interval-censored data has been previously studied \citep{Commenges2002,Cook2020}. It was found that ignoring the observation scheme of the data leads to biased estimates of regression coefficients, baseline hazards, and survival rates. A prominent motivation to illness-death model for interval-censored disease data comes from the study of dementia data \citep{frydman1995, joly2002, frydman2009, mimputation, leffondre, sabathe2020}. Dementia is diagnosed at infrequent follow-up visits which results in the time to dementia being interval-censored. Further, if a patient's last dementia test was negative and he dies, it is not known if he acquired dementia prior to death.
\cite{frydman1995} developed a non-parametric maximum likelihood procedure for the estimation of the cumulative transition hazards when times of disease are interval-censored.  Here, the second form of incompleteness was not addressed, as the author assumed that the disease state is known before death or right-censoring time. 
\cite{joly2002} proposed a non-parametric penalized likelihood method to estimate transition intensities in an illness-death model with an intermittently observed disease state. Simulations showed that not adjusting for the interval-censored nature of the data leads to a systematic bias in the estimation of transition intensities.
\cite{frydman2009} extended the \cite{frydman1995}'s methodology to incorporate the observations with unknown intermediate event status. They estimated the distribution of the time to the first occurrence of disease or death and showed that their method corrects bias.
\cite{mimputation} used multiple imputation to analyze two aspects concerning the risk of dementia: the risk of developing dementia and the impact of dementia on survival. 
\cite{leffondre} performed simulation studies to show how interval-censoring affects the estimation of the effect of risk factors.
More recently, \cite{sabathe2020} extended the pseudo-value approach \citep{andersen2003} to interval-censored data in an illness-death model by using a semi-parametric estimator based on penalized likelihood approximated by splines. This extension allowed for the direct estimation of the impact of covariates on the probability of staying alive and non-demented, on the absolute risk, and on the restricted mean survival time without dementia.

In the R software,\cite{Rsoftware} illness-death models with exactly observed event times can be estimated with several R-packages, such as the \texttt{survival} and the \texttt{mstate} packages \citep{survival-book, mstate-jss}. The number of packages that can deal with an interval-censored disease state, however, is limited. The \texttt{msm} and the \texttt{SmoothHazard} packages can fit an illness-death model for interval-censored disease times and exact death times \citep{msm-jss, sm-jss}. In the \texttt{msm} package piece-wise constant hazards need to be assumed and in the \texttt{SmoothHazard} package the user is able to choose between Weibull transition hazards and M-splines. 

While the effect of ignoring the interval-censored nature of the disease data on regression coefficients and baseline hazards has been studied \citep{Commenges2002,Cook2020}, its impact on the assessment of predictive performance has been neglected so far.
The predictive performance of a model can be evaluated using various measures of discrimination and calibration \citep{steyerberg2010}.
Model discrimination reflects how well a model can distinguish between high-risk and low-risk patients, whereas calibration assesses how closely the predicted outcomes align with the observed outcomes.
In the context of illness-death models, good calibration indicates that the predicted probabilities of transitions between the healthy and disease states are accurate and reliable. 
Good discrimination means the model can effectively identify individuals who are more likely to transition between states (e.g., from healthy to diseased, or from diseased to deceased) compared to those less likely to do so. In this article, only model discrimination is considered.

The aim of this article is to study the discrimination performance of an interval-censored binary disease marker on survival. How much does the occurrence or absence of disease contribute to survival predictions over time? The illness-death model for data in which the disease state is interval-censored is considered. 
The effect of interval-censoring on the time-specific Area Under the receiver operating characteristic Curve (AUC), for both incident/dynamic and cumulative/dynamic definition \citep{hz2005, zh2007}, is addressed by extending the standard time-dependent AUC to incorporate hazards and transition probabilities. This allows the AUC calculation to account for the uncertainty inherent in interval-censored data, which is ignored by traditional methods assuming exact times.
 Several estimation approaches are compared for two types of models: the Cox model with time-dependent disease marker and the illness-death model for interval-censored data as implemented in the \texttt{msm} and \texttt{SmoothHazard} R-packages \citep{msm-jss, sm-jss}. For this purpose a simulation study is conducted where data are simulated from an illness-death model with Weibull transition hazards and disease state censored at regular intervals. The different methods are also applied to soft tissue sarcoma data, where patients are initially disease-free after surgery, and the interval-censored disease state is the post-operative development of distant metastases.

The remainder of this article is organized as follows.
Section \ref{sec-IDM} introduces the illness-death model in general  and the different models considered in this work. Section \ref{sec:ts-auc} introduces the definitions of time-specific AUC for a binary time-dependent marker. The simulation study and the real data application are presented in Sections \ref{sec-simulation} and \ref{sec-application}, respectively. A discussion follows in Section \ref{sec-discussion}. 

\section{Illness-death models}\label{sec-IDM}
An illness-death process $\left\{V(t), t \ge 0 \right\}$ is a random process taking values in the state space $\mathcal{S} = \{0,1,2\}$ \citep{tutorial}. As shown in Figure \ref{idmodel}, subjects are initially disease-free ($V(0) = 0$) and may die directly without disease (transition $0\to 2$), or become diseased (transition $0\to 1$) and then die (transition $1\to 2$). 
Throughout the article, a non-homogeneous Markov illness-death process is assumed, meaning the future evolution of the process $\left\{V(t), t > s \right\}$ depends only on the current state $V(s)$. The transition probabilities from state $h$ to state $l$ (i.e., $h \to l$) are hence defined as follows:
\begin{equation}
	P_{hl}(s,t) = \Pr\left(V(t) = l|V(s) = h\right).
\end{equation}
If $T$ denotes the time of reaching state $l$ from state $h$, the instantaneous transition intensities are given by:
\begin{equation}
	\lambda_{hl}(t) = \lim_{\Delta t \to 0} \frac{P_{hl}(t,t+ \Delta t)}{\Delta t},
\end{equation}
with relative cumulative hazard for transition $h \to l$ defined as:
\begin{equation}
	\Lambda_{hl}(t) = \int_{0}^t \lambda_{hl}(s) ds.
\end{equation}

The disease state can also be seen as a time-dependent binary marker $X(t)$ which can take values 0 and 1, corresponding to not having and having disease at time $t$, respectively. The illness-death process at time $t$ can be represented using the time-dependent binary covariate  $X(t)$ and the time of death $T^D$, as follows:
\begin{itemize}
	\item $V(t) = 0$ corresponds to $T^D>t$ and $X(t)=0$;
	\item $V(t) = 1$ corresponds to $T^D>t$ and $X(t)=1$;
	\item $V(t) = 2$ corresponds to $T^D=t$ (for simplicity of notation, it is assumed that the subject has just died, i.e., $T^D>t-$), with two possible scenarios: (i) $X(t-)=0$ if the subject die directly, or (ii) $X(t-)=1$ if the subject first develop the disease and then die.
\end{itemize}

Often, the exact time of disease onset cannot be observed; it can only be diagnosed at pre-specified follow-up times, resulting in interval-censored data. In this article, four different methods to estimate the illness-death model data are compared: (1) the \cite{cox} model with disease state as time-dependent covariate (ignoring the interval-censored nature of the time-dependent covariate), (2) the piecewise-constant illness-death  model accounting for interval-censoring using the \texttt{msm} function from the \texttt{msm} package \citep{msm,msm-jss}, (3) the Weibull illness-death  model accounting for interval-censoring using the \texttt{idm(method = "Weib")} function from the \texttt{SmoothHazard} package, and (4) the M-spline illness-death model accounting for interval-censoring using the \texttt{idm(method = "Splines")} function from the \texttt{SmoothHazard} package \citep{smooth, sm-jss}.
All methods are now introduced and then compared in terms of discrimination performance in a simulation study (Section \ref{sec-simulation}) and in a real data application (Section \ref{sec-application}). Each model is presented without covariate adjustment, as covariates are not included in the simulation study or in the application, but they could be incorporated on a transition-specific basis to capture distinct effects.

\subsection{Cox model with time-dependent covariate}\label{sec-IDM-cox}
The Cox model with a binary time-dependent covariate is defined by the following hazard function:
\begin{equation}\label{eq:cox}
	\lambda(t | X(t)) = \lambda_0(t) \exp(\beta X(t)),
\end{equation}
where $\lambda_0(t)$ is the baseline hazard, $X(t)$ is the binary disease marker at time $t$ and $\beta$ its effect. This model can be estimated by e.g. the \texttt{coxph} R-function from the \texttt{survival} package \citep{survival-book}, which ignores the interval-censored nature of the time-dependent covariate. Disease time is assumed to be the time of diagnosis of disease: $X(t) = 0$ if a patient was not diagnosed with disease yet at time $t$ and $X(t) = 1$ if a patient was diagnosed with disease by time $t$. This approach does not account for interval censoring, a major limitation in the cases considered in this study.

The Cox model with time-dependent in \eqref{eq:cox} covariate corresponds to an illness-death multi-state model with transition hazards $\lambda_{hl}(t)$ from state $h$, ($h = 0, 1$) to state $l$, ($l = 1, 2$). The baseline hazard in Equation \eqref{eq:cox} corresponds to the transition hazard $\lambda_{02}(t)$ from state 0 (disease-free) to state 2 (death) in the multi-state model and $\lambda_0(t) \exp(\beta)$ corresponds to the transition hazard $\lambda_{12}(t)$ from state 1 (disease) to state 2 (death). The transition hazard from state 0 (disease-free) to state 1 (disease) is not formally modeled. The term $\exp(\beta)$ is the hazard ratio (HR) for disease (yes vs no).   Transition probabilities $P_{hl}(s,t)$ can be retrieved from the model using  \texttt{msfit} and \texttt{probtrans} functions from the \texttt{mstate} package \citep{mstate-jss,mstate}.

\subsection{Piecewise-constant model accounting for interval-censoring} 
Interval-censored data from an illness-death process are a special case of panel data, in which the state of an individual is observed at a finite series of times. The likelihood for panel data can be calculated in closed form if the transition hazards are constant or piece-wise constant \citep{msm-jss}. A model with piecewise-constant transition hazards $\lambda_{hl}(t)$ for the transition from state $h$, ($h = 0, 1$) to state $l$, ($l = 1, 2$) with $k$ supporting points is given by
\begin{equation}\label{eq:pwc}
	\lambda_{hl}(t) = 
	\begin{cases}
		\lambda_{hl,1} &\text{if } t \leq c_1\\
		\lambda_{hl,2} &\text{if } c_1 < t \leq c_2\\
		\vdots\\
		\lambda_{hl,k+1} &\text{if } c_{k} < t\\
	\end{cases},
\end{equation}
where $c_k$ are the times at which the hazard for transition $h \rightarrow l$ may change. This model is implemented in the \texttt{msm} package by \cite{msm-jss} and can account for interval-censoring and the probability of developing disease between last disease scan and death or lost to follow-up. 
The values $c_k$ of the piecewise-constant intervals may differ for the different transitions.

The hazards for transitions to the death state can also be assumed proportional, allowing the effect of disease on survival to be modeled by constraining the transition hazards $\lambda_{12}$ and $\lambda_{02}$ to be equal and assuming the same piecewise-constant intervals. This can be done using the \texttt{qconstrained} option and by defining a binary time-dependent covariate $X(t)$, which equals 0 if the patient has not yet been diagnosed with the disease at time $t$, and 1 if the diagnosis has occurred by that time. The transition hazard from state 1 (disease) to state 2 (death) is then equal to  
$\lambda_{12}(t) = \lambda_{02} \exp\{\beta X(t)\}$.

\subsection{Weibull model accounting for interval-censoring}
This Markov illness-death model assumes a Weibull parametrization for the transition intensities $\lambda_{hl}(t)$ from state $h$ to state $l$, as follows:
\begin{equation}\label{eq:weibull}
	\lambda_{hl}(t) =  \alpha_{hl} k_{hl} t^{k_{hl}-1}, \quad hl \in \{01,02,12\}
\end{equation}
where $\alpha_{hl}$ and $k_{hl}$ are rate and shape parameters for the transition $h \to l$, respectively. This model accounts for interval-censoring and the probability of developing disease between last disease scan and death or lost to follow-up. It is implemented in the \texttt{SmoothHazard} R-package by \cite{sm-jss} and is estimated by maximizing the likelihood with the \texttt{idm(method = "Weib")} function. 
The function does not allow for the transition hazards to the death state to be set proportional and therefore no HR of disease on death can be estimated. A time-dependent HR could be estimated by computing the ratio of $\lambda_{12}(t)$ and $\lambda_{02}(t)$. The package provides predictions of transition probabilities based on estimated transition hazards.

\subsection{M-spline model accounting for interval-censoring}
This model is estimated using a penalized likelihood approach where the three baseline transition intensities are approximated by linear combinations of M-spline basis functions, as follows:
\begin{equation}\label{eq:mspline}
	\lambda_{hl}(t) = \sum_{k=1}^{n_{hl}}(a_{hl,k})^2\,M_{hl,k}(t), \quad 
\end{equation}
where $hl \in \{01,02,12\}$ and $a_{hl,k}$ are unknown parameters.
For each transition $hl$, $M_{hl}=\left\{M_{hl,1},\dots,M_{hl,n_{hl}}\right\}$ denotes the family of $n_{hl} = m_{hl} + 2$ cubic M-splines, where $m_{hl}$ is the number of knots (7 by default). This model implemented in the \texttt{SmoothHazard} R-package, and can be estimated using the \texttt{idm} function under the option \texttt{method = "Splines"} \citep{joly2002, sm-jss}.

As the previous two models, the M-spline model accounts for interval-censoring of the disease state as well as the probability of developing disease between the last disease scan and death or lost to follow-up. 
As for the Weibull model, the transition hazards towards the death state can not be set proportional and therefore only a time-dependent HR for disease can be estimated. Transition probabilities can be obtained using functions provided in the package.

The M-spline model offers greater flexibility than parametric models, but it also requires careful handling of the smoothing parameters in the penalized likelihood to ensure stable estimation. To this purpose, the \texttt{idm} function provides the option to automatically select the smoothing parameters using approximate leave-one-out cross-validation (\texttt{CV=1}) \citep{sm-jss} However, cross-validation can be computationally intensive, and manual tuning of the smoothing parameters may be preferred, which can be challenging in complex datasets and is not always suitable for simulation studies.

\section{Discrimination for time-dependent marker and survival data}\label{sec:ts-auc}
Discrimination of a time-dependent marker in survival analysis refers to the ability of the marker to distinguish  between individuals who will experience an event and those who will not at different time points. 
Several measures of discrimination performance have been introduced in the field of survival analysis. In this article, discrimination is assessed using the time-specific Area Under the receiver operating characteristic Curve (AUC), defined according to sensitivity and specificity for survival outcome and a longitudinal binary marker.

Originally, sensitivity and specificity were introduced in the context of evaluating a time-fixed marker $X$ used to classify a binary outcome $Y$ for \textit{cases} ($Y=1$) versus \textit{controls} ($Y=0$), via a threshold $c$. A subject is predicted to be a case if the value of the covariate is bigger than $c$; otherwise it is predicted to be a control. Sensitivity and specificity are hence defined as follows:
\begin{align*}
	\text{sensitivity}(c) &= \text{true positive rate} = \Pr(X>c|Y=1);\\ \text{specificity}(c) &= \text{true negative rate} = \Pr(X \leq c|Y=0).
\end{align*}
The full range of sensitivity and specificity for different classification criteria $c$ can be graphically summarized by the receiver operating characteristic (ROC) curve, which plots sensitivity against 1-specificity \citep{fawcett}.  The ROC curve reflects differences in the marker distribution between cases and controls. If the distributions are identical (i.e., the marker is useless to distinguish cases from controls), the curve follows the 45-degree line and the AUC equals 0.5, indicating no discrimination. The AUC measures how well the marker 
$X$ distinguishes between outcomes and corresponds to the probability that the classifier will rank a randomly chosen positive instance higher than a randomly chosen negative instance \citep{fawcett}. The definition of AUC that accounts for ties  \citep{muschelli2020} is:
\begin{equation*}
	\text{AUC} = \Pr\left(X_i>X_j \mid Y_i = 1, Y_j=0\right) + 0.5 \cdot \Pr\left(X_i=X_j \mid Y_i = 1, Y_j=0\right),
\end{equation*}
where $i$ and $j$ are randomly chosen case and control, respectively, with $\left(X_i,Y_i\right)\perp\!\!\!\perp \left(X_j,Y_j\right)$.

To extend the concepts of sensitivity and specificity to settings with censored data and longitudinal markers, several definitions of cases and controls have been proposed. In this article, two of these definitions are considered: 1) incident cases and dynamic controls, and 2) cumulative cases and dynamic controls \citep{heagerty2000, hz2005, zh2007}.
The two approaches target distinct aspects of marker performance: incident/dynamic metrics evaluate how well the marker discriminates individuals who will experience death at a specific time point among those still at risk, whereas cumulative/dynamic metrics assess how well the marker discriminates individuals who will experience death within a given time horizon.

\subsection{Incident cases and dynamic controls}\label{sec:ts-auc:id}
\cite{hz2005} defined incident sensitivity and dynamic specificity at time $t$ as follows
\begin{align*}
	\text{sensitivity}^{I}(c ,t) &= \Pr(X(t) > c \mid T^D = t)\\
	\text{specificity}^{D}(c ,t) &= \Pr(X(t) \leq c \mid T^D > t),
\end{align*}
where $c$ is a classification criterion, $T^D$ is time of death, and $X(t)$ is the time-dependent disease marker evaluated at time $t$. In this definition the individuals who die at time $t$ are considered \textit{cases} and individuals who survive beyond time $t$ are considered \textit{controls}. Let $i, j$ be individuals, $X_i(t), X_j(t)$ their marker values at time $t$, and $T^D_i$ and $T^D_j$ their death times. The incident/dynamic AUC is then defined by \cite{hz2005}
\begin{align}
	\text{AUC}^{I/D}(t) = &\Pr(X_i(t) > X_j(t) \mid T^D_i = t, T^D_j > t)\nonumber\\
	&+ 0.5 \Pr(X_i(t) = X_j(t) \mid T^D_i = t, T^D_j > t).
\end{align}

In case $X_i(t)$ and $X_j(t)$ are binary covariates the $\text{AUC}^{I/D}(t)$ can be rewritten as
\begin{equation}\label{aucid}
	\text{AUC}^{I/D}(t) = 0.5  + 0.5(p(t) -\pi_1(t))
\end{equation}
where $\pi_{1}(t)$ is the probability that a person alive at time $t$ has experienced disease (prevalence of disease), and $p(t)$ is the probability that a person who dies at time $t$ has a history of disease (see the demonstration in Supplementary Material A.1).  
As mentioned in Section \ref{sec-IDM}, the disease marker $X(t)$ is related to the illness-death model of Figure \ref{idmodel} in the following way: $X(t) = 0$ if a patient did not move to state 1 (disease) before time $t$ (in state 0 or 2 at time $t$) and $X(t) = 1$ if a patient moved to state 1 (disease) before time $t$ (in state 1 or 2 at time $t$).
The terms $\pi_{1}(t)$ and $p(t)$ can be expressed by transition probabilities in a illness-death model, as follows:
\begin{align}
	\pi_{1}(t) = & \Pr\left(X_i(t)=1 \mid T^D_i > t\right)
	=  \frac{P_{01}(t)}{P_{00}(t) + P_{01}(t)} \label{pi-id}, \quad \text{ and }\\
	p(t) = & \Pr\left(X_i(t-)=1 \mid T^D_i = t\right)
	=  \frac{P_{01}(t-)\lambda_{12}(t)}{P_{00}(t-)\lambda_{02}(t) + P_{01}(t-)\lambda_{12}(t)},\label{p-id}
\end{align}
where $t-$ means just before time $t$, $P_{0l}(t)$ is the conditional probability of being in state $l = 0, 1$ at time $t$ given state 0 at time 0, and $\lambda_{h2}(t)$ is the transition hazard at time $t$ for moving from state $h = 0, 1$ to state 2. 
Equations \eqref{aucid}--\eqref{p-id} therefore relate the incident/dynamic AUC to transition probabilities and hazards.

The incident/dynamic AUC at a specific time $t$ measures how well the disease marker evaluated at time $t$ separates those who die at $t$ from those who survive.
By considering $\gamma(t) = \lambda_{12}(t)/\lambda_{02}(t)$, the difference between $p(t)$ and $\pi_{1}(t)$ is equal to 
\begin{align}
	p(t) - \pi_{1}(t) = & \frac{\gamma(t)P_{01}(t)}{P_{00}(t) + \gamma(t)P_{01}(t)} - \frac{P_{01}(t)}{P_{00}(t) + P_{01}(t)} \nonumber\\
	&= (\gamma(t) - 1) \frac{P_{01}(t)P_{00}(t)}{(P_{00}(t) + \gamma(t)P_{01}(t))(P_{00}(t) + P_{01}(t))}\label{ppi}\\
	&= (\gamma(t) - 1) \frac{1}{(1 + \gamma(t)P_{01}(t)/P_{00}(t))(1 + P_{00}(t)/P_{01}(t))}.\nonumber
\end{align}
From \eqref{aucid} and \eqref{ppi} follows that if $\gamma(t) \equiv 1$ then $\text{AUC}^{I/D}(t) = 0.5$, if $\gamma(t) > 1$ then $\text{AUC}^{I/D}(t) \geq 0.5$ and if $\gamma(t) < 1$ then $\text{AUC}^{I/D}(t) \leq 0.5$.

\subsubsection{Estimation.}
Estimates for the incident/dynamic AUC can be obtained by replacing transition probabilities and hazards by their estimated counterparts: 
\begin{equation}
	\widehat{\text{AUC}}^{I/D}(t) = 0.5  + 0.5\left(\hat{p}(t) -\hat\pi_1(t)\right),\label{eq-estimate-id}
\end{equation}
where
$$\hat{\pi}_{1}(t) =  \frac{\hat{P}_{01}(t)}{\hat{P}_{00}(t) + \hat{P}_{01}(t)} \quad \text{ and } \quad
\hat{p}(t) = \frac{\hat{P}_{01}(t-)\frac{\hat{\lambda}_{12}(t)}{\hat{\lambda}_{02}(t)}}{\hat{P}_{00}(t-) + \hat{P}_{01}(t-)\frac{\hat{\lambda}_{12}(t)}{\hat{\lambda}_{02}(t)}},$$
and $t-$ means just before time $t$, $\hat{P}_{0l}(t)$ is an estimate of the conditional probability of being in state $h$ at time $t$ given $V(0)=0$, and $\hat{\lambda}_{h2}(t)$ is an estimate of the transition hazard at time $t$ for moving from state $h$ to state 2. 
Such estimates may be obtained from software packages for multi-state models, such as the R-packages \texttt{mstate} \citep{mstate-jss}, \texttt{msm} \citep{msm-jss}, and \texttt{SmoothHazard} \citep{sm-jss} discussed in Section \ref{sec-IDM}. For the Cox model with time-dependent covariate (Section \ref{sec-IDM-cox}), the incident/dynamic AUC can also be estimated using the \texttt{risksetAUC} function from the \texttt{risksetROC} R-package by \cite{risksetROC}.

\subsection{Cumulative cases and dynamic controls}\label{sec:ts-auc:cd}
\cite{zh2007} defined the cumulative sensitivity and dynamic specificity at time $t$ for a time-dependent covariate evaluated at time $s < t$ as
\begin{align*}
	\text{sensitivity}^{C}(c \mid \text{start} = s, \text{stop} = t) &= \Pr(X(s) > c \mid s < T^D\leq t)\\
	\text{specificity}^{D}(c \mid \text{start} = s, \text{stop} = t) &= \Pr(X(s) \leq c \mid T^D > t),
\end{align*}
where $T^D$ is time of death, $X(s)$ is marker measurement at time $s$.
\textit{Cases} are individuals who die within a time window ($t-s$) from $s$ and \textit{controls} are individuals who survive the time window.
The cumulative/dynamic AUC is then defined by
\begin{align}
	\text{AUC}^{C/D}(s, t) =  &\Pr\left(X_i(s) > X_j(s) \mid s < T^D_i \le t, T^D_j > t\right)\nonumber\\
	&+ 0.5 \Pr\left(X_i(s) = X_j(s) \mid s < T^D_i \le t, T^D_j > t\right),
\end{align}
where $i, j$ are individuals, $X_i(s), X_j(s)$ their marker values at time $s$, and  $T^D_i, T^D_j$ their death times. For binary $X_i(s)$ and $X_j(s)$ the $\text{AUC}^{C/D}$ can be rewritten as
\begin{equation}\label{auccd}
	\text{AUC}^{C/D}(s, t) = 0.5 + 0.5(p(s,t) - \pi_{1}(s, t)),
\end{equation}
where $\pi_{1}(s, t)$ is the probability that a person alive at time $t$ had experienced disease by time $s$ and $p(s,t)$ is the probability that a person who dies in the time interval $(s, t]$ had experienced disease by time $s$ (see the demonstration in Supplementary Material A.2). The quantities $\pi_{1}(s, t)$ and $p(s,t)$ can be written in terms of transition probabilities:
\begin{equation}\label{pi-cd}
	\pi_{1}(s, t) = \Pr(X_j(s)=1 \mid T^D_j > t) =  \frac{P_{01}(0,s)P_{11}(s,t)}{P_{00}(0,t) + P_{01}(0,t)}\\
\end{equation}
and
\begin{align}
	p(s,t) &= \Pr(X_i(s)=1 \mid T^D_i > s, T^D_i \leq t) \nonumber \\ &=  \frac{P_{01}(0,s)P_{12}(s,t)}{P_{00}(0,s)P_{02}(s,t) + P_{01}(0,s)P_{12}(s,t)}, \label{p-cd}
\end{align}
where $P_{kl}(u, v)$ is the conditional probability of being in state $l$ at time $v$ given in state $h$ at time $u$. 
Equations \eqref{auccd}--\eqref{p-cd} hence relate the cumulative/dynamic AUC to transition probabilities.

The cumulative/dynamic AUC at time $s$ measures how well the disease marker evaluated at time $s$ separates those who die before time $t$ from those who survive until $t$.

\subsubsection{Estimation.} Estimates for the cumulative/dynamic AUC can be obtained by replacing transition probabilities and hazards by their estimated counterparts:
\begin{equation}
	\widehat{\text{AUC}}^{C/D}(s, t) = 0.5 + 0.5(\hat{p}(s,t) - \hat{\pi}_{1}(s, t)),
\end{equation}
where
$$ \hat{\pi}_{1}(s, t) = \frac{\hat{P}_{01}(0,s)\hat{P}_{11}(s,t)}{\hat{P}_{00}(0,t) + \hat{P}_{01}(0,t)}
\quad \text{ and } \quad
\hat{p}(s,t) = \frac{\hat{P}_{01}(0,s)\hat{P}_{12}(s,t)}{\hat{P}_{00}(0,s) \hat{P}_{02}(s,t)+ \hat{P}_{01}(0,s)\hat{P}_{12}(s,t)},$$
and $\hat{P}_{hl}(u, v)$ is an estimate of the conditional probability of being in state $l$ at time $v$ given in state $h$ at time $u$. Such estimates may be obtained from software packages for multi-state models, such as the R-packages \texttt{mstate} \citep{mstate-jss}, \texttt{msm} \citep{msm-jss}, and \texttt{SmoothHazard} \citep{sm-jss}  discussed in Section \ref{sec-IDM}.

\section{Simulation study}\label{sec-simulation}
To study the discrimination performance of an interval-censored binary disease marker on survival a simulation study was conducted. Incident/dynamic and cumulative/dynamic AUC were computed for the different estimation procedures presented in Section \ref{sec-IDM}: the Cox model with time-dependent disease marker, which ignores interval-censoring, and the illness-death model with piecewise-constant, Weibull or M-spline transition hazards.
For the piecewise-constant model, the four change points $c_k$  at which the hazard may change were considered at 6, 30, 60, and 90 months.
For the M-spline model, the default of 7 knots per transition was used. 
Automatic selection of the smoothing parameter via approximate leave-one-out cross-validation was not implemented, as this approach may be numerically unstable in simulation studies;  similarly, manual tuning was avoided to ensure consistency and comparability across simulated datasets.
Multiple data scenarios were simulated (Section \ref{sec-simu-data}) and results from the different methods were compared (Section \ref{sec-simu-res}). R source code is available at \url{http://github.com/mspreafico/auc-idmIC}.

\subsection{Data generation and methods}\label{sec-simu-data}
Data were generated from Weibull transition hazards with a common shape parameter $k$ and different rate parameters $\alpha_{01}$, $\alpha_{02}$, and $\alpha_{12}$, using the following procedure. Let $y$ denote the total follow-up length, with visits occurring at every $\tau$ time units, so that the time-dependent binary covariate $X(t)$ for illness can be observed at visits $t = 0,\tau,2\tau,3\tau,\dots,y$. For each individual:

\begin{enumerate}
	\item Draw $U_1 \sim \text{Uniform}(0,1)$ and compute the Weibull-distributed first true exit time from state $h=0$ (disease-free) using the inversion method: $$T_{1} = \left(\frac{-\ln U_1}{\alpha_{01} + \alpha_{02}}\right)^{1/k}.$$
	
	\item Decide which transition $h \rightarrow l$ occurred at $T_1$ by randomly sampling the first exit state $l \in \{1,2\}$ with probabilities
	$$P(l=1) = \frac{\alpha_{01}}{\alpha_{01} + \alpha_{02}}, \quad
	P(l=2) = \frac{\alpha_{02}}{\alpha_{01} + \alpha_{02}}.$$
	
	If $l=2$, then the individual died at time $T^D = T_1$ without illness. \\
	If $l=1$, then the individual enters the disease state and $T_1$ is true illness time.
	
	\item If the individual entered the disease state, draw $U_2 \sim \text{Uniform}(0,1)$ and compute the true death time using the inversion method conditional on having already waited to $T_1$:
	\[
	T^D = \left(T_1^k - \frac{\ln U_2}{\alpha_{12}}\right)^{1/k}.
	\]
	
	\item Censor the true death time according to the desired censoring scheme:
	\begin{itemize}
		\item \textit{Administrative censoring:} the survival time is $T^* = \min(T^D, y)$ with death indicator $\delta = \mathbf{1}\{T^D \leq y\}$;
		\item \textit{Uniform random censoring:} draw the censoring time $C \sim \text{Uniform}(a,y)$ and determine the survival time as $T^* = \min(T^D, C)$ with death indicator $\delta = \mathbf{1}\{T^D \leq C\}$.    
	\end{itemize}
	
	\item Let $v^* = \max\{t\le T^*: t\in\{0,\tau,2\tau,\dots,y\}\}$ denote the last observed visit before death or censoring. If the individual died without illness or was censored at $T^* < T_1$, then $X(t)=0$ for all $t=0,\tau,2\tau,\dots, v^*$. For individuals that entered the disease state before dying or being censored ($T^* \le T^1$), map the true illness time $T_1$ to the first scheduled visit at/after illness and determine:  
	$$X(t) = \begin{cases}
		0 \quad \text{ for } t=0,\dots,v,\\
		1 \quad \text{ for } t=v,\dots,v^*,\\
	\end{cases}$$
	where  $v = \min\{t\ge T_1: t\in\{0,\tau,2\tau,\dots,v^*\}\}$
	denotes the first visit after $T_1$, i.e., the observed illness time. In both cases $X(T^*) = X(v^*)$.
	
	\item Return the time-to-death outcome $(T^*, \delta)$, and the time-dependent covariate $\left\{X(t): t\in \{0,\tau,\dots,v^*,T^*\}\right\}$ for observed disease.
	
\end{enumerate}

Table \ref{tab-scenarios} summarizes the 18 scenarios (A--R) generated using shape parameter $k = 0.5$ and rate parameters $\alpha_{01} = \alpha_{02} = 0.05$ and $\alpha_{12} = 0.56$. These Weibull parameters were kept fixed throughout the simulated scenarios and were based on the data discussed in Section \ref{sec-application}, assuming a proportional hazards between diseased and non diseased with HR equal to 11.2. 
The number of individuals per data set, $N$, was either 400, 1000 or 2000.  
The survival time was censored according to two different censoring schemes: either it was censored administratively at 10 years ($y = 120$ months) follow-up or censoring times were sampled from a uniform distribution between 5 and 10 years ($a=60$ and $y = 120$ months).
The disease state was observed only at pre-specified follow-up visits. The scenarios cover three different follow-up schemes in which the disease state was observed every 3, 6, or 12 months ($\tau = 3, 6, 12$).

The true incident/dynamic and cumulative/dynamic AUC values over time were determined using Equations \eqref{aucid} and \eqref{auccd}, where the terms $\pi_{1}(t)$ and $p(t)$ were computed from the transition probabilities and hazards derived from the Weibull illness-death model with $k = 0.5$, $\alpha_{01} = \alpha_{02} = 0.05$, and $\alpha_{12} = 0.56$, as detailed in Supplementary Material B.
Each scenario was simulated $n_{sim}=1000$ times. 
For each model and scenario, estimated incident/dynamic and cumulative/dynamic AUCs over the years were presented graphically.
Estimated bias, empirical standard error, and root mean squared error \citep{Morris2019} of both AUCs at specific time-points were considered as performance measures.

\begin{table}[t]
	\small\centering
	\caption{Simulated scenarios. $N$ is the total number of patients. Censoring is the type of death censoring: Unif(5, 10) means censoring was uniformly sampled between 5 and 10 years and 10 means that administrative censoring occurred at 10 years. Follow-up is time between disease observations in months.\label{tab-scenarios}}
	\begin{tabular}{cccr}
		\toprule
		Scenario & $N$ & Censoring & Follow-up \\ 
		\midrule
		A & 1000 & Unif(5, 10) & 3 \\ 
		B &  1000 & Unif(5, 10) & 6 \\ 
		C &  1000 & Unif(5, 10) & 12 \\ 
		D &  1000 & 10 & 3 \\ 
		E &  1000 & 10 & 6 \\ 
		F &  1000 & 10 & 12 \\ 
		G &  2000 & Unif(5, 10) & 3 \\ 
		H &  2000 & Unif(5, 10) & 6 \\ 
		I &  2000 & Unif(5, 10) & 12 \\ 
		J &  2000 & 10 & 3 \\ 
		K &  2000 & 10 & 6 \\ 
		L &  2000 & 10 & 12 \\
		M &  400 & Unif(5, 10) & 3 \\ 
		N &  400 & Unif(5, 10) & 6 \\ 
		O &  400 & Unif(5, 10) & 12 \\ 
		P &  400 & 10 & 3 \\ 
		Q &  400 & 10 & 6 \\ 
		R &  400 & 10 & 12 \\
		\bottomrule
	\end{tabular}
\end{table}

\subsection{Results}\label{sec-simu-res}

\subsubsection{Estimated models.}
Hazard ratios for disease (yes vs no) were estimated for the piecewise-constant and the Cox model. Average values over $n_{sim}=1000$ repetitions were summarized in Supplementary Table C.1. For the Weibull and M-spline model no effect could be estimated, since the \texttt{idm} function does not allow transition hazards to be proportional. The coefficients from the Cox model were consistently more biased than the ones from the piecewise-constant model. 
The Cox model underestimated the true coefficient and the bias increased for larger follow-up intervals. 
Empirical SE decreased as sample size increased and was comparable between the Cox and piecewise-constant models, as well as across scenarios with equal sample sizes.
These results are in line with \cite{leffondre} who showed that the effect estimates of the Cox model were biased if the covariate affected both the risk of disease and death.

The M-spline model did not converge for many simulated data sets, with the frequency of invalid estimations (Supplementary Table C.2) increasing as sample size decreased. These difficulties are partly attributable to the decision not to implement automatic smoothing parameter selection due to potential numerical instability in simulations and to avoid manual tuning to ensure comparability across datasets.
These invalid estimates prevented the estimation of the incident/dynamic and cumulative/dynamic AUCs. Therefore, the results for the M-spline model shown in Sections \ref{sec-simu-res:id} and \ref{sec-simu-res:cd} are only based on the obtained valid estimates.

Since average AUC estimates were nearly identical between scenarios where only the sample size differed, only results for scenarios A to F  with $N = 1000$ are shown below.  Results of the other scenarios are shown in Supplementary Material C.

\subsubsection{Incident/dynamic AUC.}\label{sec-simu-res:id}
For each model and scenario A-F, Table \ref{tab-auc-id} shows the bias, empirical Standard Error (SE), and Root Mean Squared Error (RMSE) for estimates of the incident/dynamic AUC at years 1, 3, and 5, which coincide with the times of follow-up visits for every scenario. 
The Weibull model outperformed the other models in each scenario. This is not surprising since the data were generated according to Weibull distributions.
The M-spline model consistently had the largest empirical standard error as well as the second smallest bias overall.
The piecewise-constant model was slightly less biased than the Cox model for scenarios with 6 and 12 months between follow-up visits (scenarios B, C, E, F). For the scenarios with 3 months in between follow-up visits, the Cox model outperformed the piecewise-constant model (scenarios A, D) in terms of bias. 
The follow-up schemes with larger intervals resulted in larger bias of the incident/dynamic AUC estimates, particularly for the Cox model.
Although the censoring scheme impacted the valid estimates of the M-spline model (Supplementary Table C.2), it did not have a major effect on the AUC estimates for the Cox, piecewise-constant, and Weibull models.

\begin{table}[h!]
\centering
	\caption{Estimated bias, empirical standard error (SE), and root mean square error (RMSE) for time-specific incident/dynamic AUC at 1, 3, 5 years under scenarios A to F in Table \ref{tab-scenarios}.}\label{tab-auc-id}
	\begin{tabular}{clrrrrrrrrr}
		\toprule 
		&  & \multicolumn{3}{c}{\underline{$\text{AUC}^{\text{I/D}}$(1) = 0.71}} & \multicolumn{3}{c}{\underline{$\text{AUC}^{\text{I/D}}$(3) = 0.72}} & \multicolumn{3}{c}{\underline{$\text{AUC}^{\text{I/D}}$(5) = 0.72}} \\
		Scenario & Model & Bias & SE & RMSE & Bias & SE & RMSE & Bias & SE & RMSE \\ 
		\midrule
		A & Cox & -0.04 & 0.01 & 0.04 & -0.02 & 0.01 & 0.02 & -0.02 & 0.01 & 0.02 \\ 
		& PW-const & -0.05 & 0.01 & 0.05 & -0.03 & 0.01 & 0.03 & -0.02 & 0.01 & 0.03 \\ 
		& Weibull & 0.00 & 0.01 & 0.01 & 0.00 & 0.01 & 0.01 & 0.00 & 0.01 & 0.01 \\ 
		& M-spline & 0.02 & 0.02 & 0.03 & 0.00 & 0.02 & 0.02 & -0.01 & 0.04 & 0.04 \\ 
		\midrule
		B & Cox & -0.07 & 0.02 & 0.08 & -0.04 & 0.01 & 0.04 & -0.03 & 0.01 & 0.03 \\ 
		& PW-const & -0.06 & 0.01 & 0.07 & -0.04 & 0.01 & 0.04 & -0.03 & 0.01 & 0.03 \\ 
		& Weibull & -0.01 & 0.02 & 0.02 & 0.00 & 0.01 & 0.01 & 0.00 & 0.01 & 0.01 \\ 
		& M-spline & 0.01 & 0.03 & 0.03 & -0.01 & 0.03 & 0.03 & 0.00 & 0.04 & 0.04 \\
		\midrule
		C & Cox & -0.21 & 0.00 & 0.21 & -0.07 & 0.01 & 0.07 & -0.06 & 0.01 & 0.06 \\ 
		& PW-const & -0.09 & 0.01 & 0.09 & -0.06 & 0.01 & 0.06 & -0.05 & 0.02 & 0.05 \\ 
		& Weibull & -0.01 & 0.03 & 0.03 & 0.00 & 0.02 & 0.02 & 0.00 & 0.02 & 0.02 \\ 
		& M-spline & -0.05 & 0.04 & 0.06 & 0.00 & 0.03 & 0.03 & 0.00 & 0.04 & 0.04 \\ 
		\midrule
		D & Cox & -0.04 & 0.01 & 0.04 & -0.02 & 0.01 & 0.02 & -0.02 & 0.01 & 0.02 \\ 
		& PW-const & -0.05 & 0.01 & 0.05 & -0.03 & 0.01 & 0.03 & -0.02 & 0.01 & 0.03 \\ 
		& Weibull & 0.00 & 0.01 & 0.01 & 0.00 & 0.01 & 0.01 & 0.00 & 0.01 & 0.01 \\ 
		& M-spline & 0.02 & 0.02 & 0.03 & -0.01 & 0.02 & 0.03 & 0.00 & 0.03 & 0.04 \\ 
		\midrule
		E & Cox & -0.07 & 0.02 & 0.07 & -0.04 & 0.01 & 0.04 & -0.03 & 0.01 & 0.03 \\ 
		& PW-const & -0.06 & 0.01 & 0.06 & -0.04 & 0.01 & 0.04 & -0.03 & 0.01 & 0.03 \\ 
		& Weibull & -0.01 & 0.02 & 0.02 & 0.00 & 0.01 & 0.01 & 0.00 & 0.01 & 0.01 \\ 
		& M-spline & 0.01 & 0.02 & 0.03 & -0.01 & 0.03 & 0.03 & 0.00 & 0.04 & 0.04 \\ 
		\midrule
		F & Cox & -0.21 & 0.00 & 0.21 & -0.07 & 0.01 & 0.07 & -0.06 & 0.01 & 0.06 \\ 
		& PW-const & -0.09 & 0.01 & 0.09 & -0.06 & 0.01 & 0.06 & -0.05 & 0.01 & 0.05 \\ 
		& Weibull & -0.01 & 0.02 & 0.02 & 0.00 & 0.01 & 0.01 & 0.00 & 0.01 & 0.01 \\ 
		& M-spline & -0.05 & 0.04 & 0.06 & 0.00 & 0.03 & 0.03 & 0.00 & 0.04 & 0.04 \\ 
		\bottomrule
		\multicolumn{11}{p{14cm}}{\footnotesize \textbf{Abbreviations:} $\text{AUC}^{\text{I/D}}(t)$, incident/dynamic AUC at year $t$; PW-const, piecewise-constant model; Cox estimates are based on transition probabilities of Cox model through \texttt{mstate}.}
	\end{tabular}
\end{table}

\begin{figure}[h!]
	\centering
	\includegraphics[width=0.85\textwidth]{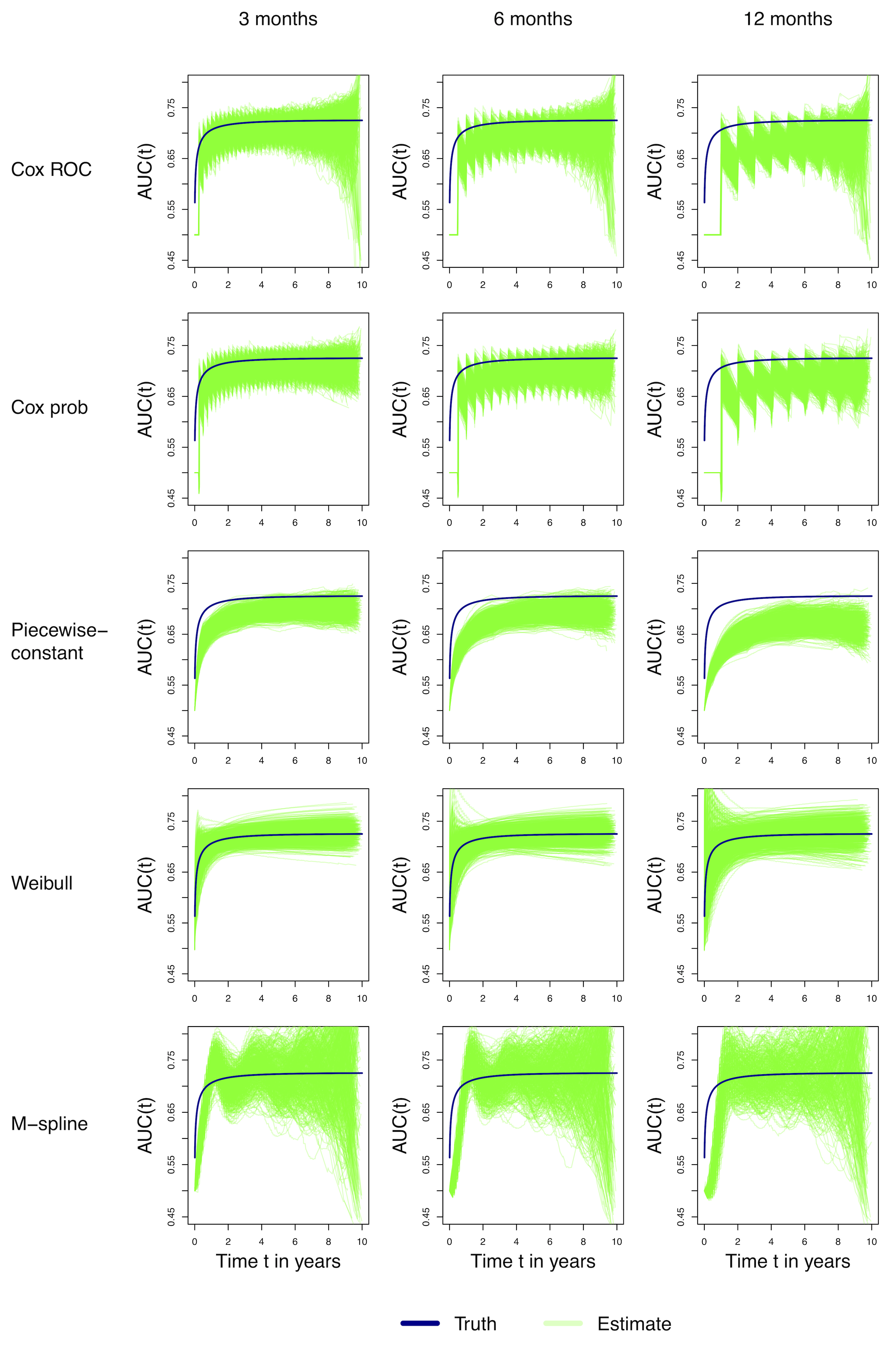}
	\caption{Estimated time-specific  incident/dynamic AUC for scenario A (3 months; left panels), B (6 months; middle panels) and C (12 months; right panels) using different models (Cox, PW-const, Weibull, M-spline). The x-axis represents time $t$ in years; the y-axis represents $\widehat{\text{AUC}}^{\text{I/D}}(t)$. The blue line in each panel represents the true values over time.
		Estimates for \textit{Cox ROC} are based on \texttt{risksetAUC} function for Cox model. Estimates for \textit{Cox prob} are based on transition probabilities of Cox model via \texttt{mstate} package.}
	\label{fig-auc-id-abc}
\end{figure}

Figure \ref{fig-auc-id-abc} shows the estimated incident/dynamic AUC over years, i.e., $\widehat{\text{AUC}}^{\text{I/D}}(t)$,  for scenarios A, B and C with follow-up visits every 3, 6, and 12 months, respectively (see Supplementary Figure C.1 for scenarios D-E-F). Each plot depicts the true AUC in blue and the estimated AUC for each simulated data set in green. 
The incident/dynamic AUC for the Cox model was estimated by two different approaches. The first approach computes the AUC from the ROC curve derived from the estimated sensitivity and specificity and is implemented in the \texttt{risksetAUC} function from the \texttt{risksetROC} R-package \citep{hz2005}. The second approach computes the AUC from estimated transition probabilities as described in Equation \eqref{eq-estimate-id}. Both methods are depicted in Figure \ref{fig-auc-id-abc}; however since the two estimation procedures for the Cox model's AUC gave similar result, only results for the transition probability based AUC are presented in Table \ref{tab-auc-id} (see Supplementary Table C.3 for the other results).
The Cox model's incident/dynamic AUC exhibits jumps at the observation time points. This occurs because, at those time points, the proportion of diseased individuals increases in the risk set. As a result, there is less bias at the times of follow-up visits compared to the times between follow-up visits. Before the first observation time point, the curve is equal to 0.5 because no disease has been observed yet.
The incident/dynamic AUC based on the M-spline model shows a similar behaviour to that of the Cox model. No distinct jumps are observed but waves can be seen that are most defined at the beginning of follow-up time. 
Since the piecewise-constant model and the Weibull model make assumptions about the hazard function, the AUC estimates do not display jumps or waves, like for the Cox and M-spline model. 
Results indicate that the piecewise-constant model is not flexible enough to follow the shape of the true AUC curve.

\subsubsection{Cumulative/dynamic AUC.}\label{sec-simu-res:cd}
Table \ref{tab-auc-cd} shows the bias, empirical Standard Error (SE), and Root Mean Squared Error (RMSE) for estimates of the incident/dynamic AUC at years 1, 3, and 5 for the four models in scenarios A to F.
The piecewise-constant model showed the worst performance and underestimated the true AUC. The Weibull model, M-spline and the Cox model provided good results.
The follow-up scheme with larger intervals resulted in consistently more biased estimates for the piecewise-constant model. The Cox, Weibull and M-spline model based estimates were of limited bias for the different follow-up schemes. 
Again, although the censoring scheme impacted the valid estimates of the M-spline model (Supplementary Table C.2), it did not have a large effect on the AUC estimates for the other models.

\begin{table}[h!]
\centering
	\caption{Estimated bias, empirical standard error (SE), and root mean square error (RMSE) for time-specific cumulative/dynamic AUC for prediction time 1, 3, 5 years and prediction window of 5 years under scenarios A to F in Table \ref{tab-scenarios}.}\label{tab-auc-cd}
	\begin{tabular}{clrrrrrrrrr}
		\toprule
		& &  \multicolumn{3}{c}{\underline{$\text{AUC}^{\text{C/D}}$(1,6) = 0.59}} & \multicolumn{3}{c}{\underline{$\text{AUC}^{\text{C/D}}$(3,7) = 0.62}} & \multicolumn{3}{c}{\underline{$\text{AUC}^{\text{C/D}}$(5,8) = 0.64}}\\
		Scenario & Model & Bias & SE & RMSE & Bias & SE & RMSE & Bias & SE & RMSE \\ 
		\midrule
		A & Cox & 0.00 & 0.01 & 0.01 & 0.00 & 0.01 & 0.01 & 0.00 & 0.02 & 0.02 \\ 
		& PW-const & -0.02 & 0.01 & 0.02 & -0.01 & 0.01 & 0.02 & -0.01 & 0.01 & 0.02 \\ 
		& Weibull & 0.00 & 0.01 & 0.01 & 0.00 & 0.01 & 0.01 & 0.00 & 0.01 & 0.01 \\ 
		& M-spline & 0.00 & 0.01 & 0.01 & 0.00 & 0.01 & 0.01 & 0.00 & 0.02 & 0.02 \\ 
		\midrule
		B & Cox & 0.00 & 0.01 & 0.01 & 0.00 & 0.01 & 0.01 & 0.00 & 0.02 & 0.02 \\ 
		& PW-const & -0.04 & 0.00 & 0.04 & -0.02 & 0.01 & 0.03 & -0.02 & 0.01 & 0.02 \\ 
		& Weibull & 0.00 & 0.01 & 0.01 & 0.00 & 0.01 & 0.01 & 0.00 & 0.01 & 0.01 \\ 
		& M-spline & 0.00 & 0.01 & 0.01 & 0.00 & 0.01 & 0.01 & 0.00 & 0.02 & 0.02 \\ 
		\midrule
		C & Cox & 0.00 & 0.01 & 0.01 & 0.00 & 0.01 & 0.01 & 0.00 & 0.02 & 0.02 \\ 
		& PW-const & -0.05 & 0.00 & 0.05 & -0.04 & 0.01 & 0.04 & -0.03 & 0.01 & 0.03 \\ 
		& Weibull & -0.01 & 0.01 & 0.01 & 0.00 & 0.01 & 0.01 & 0.00 & 0.01 & 0.01 \\ 
		& M-spline & 0.00 & 0.01 & 0.01 & 0.00 & 0.01 & 0.01 & 0.00 & 0.02 & 0.02 \\
		\midrule
		D & Cox & 0.00 & 0.01 & 0.01 & 0.00 & 0.01 & 0.01 & 0.00 & 0.01 & 0.01 \\ 
		& PW-const & -0.02 & 0.01 & 0.02 & -0.01 & 0.01 & 0.02 & -0.01 & 0.01 & 0.02 \\ 
		& Weibull & 0.00 & 0.01 & 0.01 & 0.00 & 0.01 & 0.01 & 0.00 & 0.01 & 0.01 \\ 
		& M-spline & 0.00 & 0.01 & 0.01 & 0.00 & 0.01 & 0.01 & 0.00 & 0.02 & 0.02 \\ 
		\midrule
		E & Cox & 0.00 & 0.01 & 0.01 & 0.00 & 0.01 & 0.01 & 0.00 & 0.01 & 0.01 \\ 
		& PW-const & -0.04 & 0.00 & 0.04 & -0.03 & 0.01 & 0.03 & -0.02 & 0.01 & 0.02 \\ 
		& Weibull & 0.00 & 0.01 & 0.01 & 0.00 & 0.01 & 0.01 & 0.00 & 0.01 & 0.01 \\ 
		& M-spline & 0.00 & 0.01 & 0.01 & 0.00 & 0.01 & 0.01 & 0.00 & 0.02 & 0.02 \\
		\midrule
		F & Cox & 0.00 & 0.01 & 0.01 & 0.00 & 0.01 & 0.01 & 0.00 & 0.01 & 0.01 \\ 
		& PW-const & -0.05 & 0.00 & 0.05 & -0.04 & 0.01 & 0.04 & -0.03 & 0.01 & 0.03 \\ 
		& Weibull & -0.01 & 0.01 & 0.01 & 0.00 & 0.01 & 0.01 & 0.00 & 0.01 & 0.01 \\ 
		& M-spline & 0.00 & 0.01 & 0.01 & 0.00 & 0.01 & 0.01 & 0.00 & 0.02 & 0.02 \\ 
		\bottomrule
		\multicolumn{11}{p{14cm}}{\footnotesize \textbf{Abbreviations:} $\text{AUC}^{C/D}(t,t+5)$, cumulative/dynamic AUC for prediction time $t$ in years and prediction window of 5 years}\\
	\end{tabular}
\end{table}

\begin{figure}[h!]
	\centering
	\includegraphics[width=0.85\textwidth]{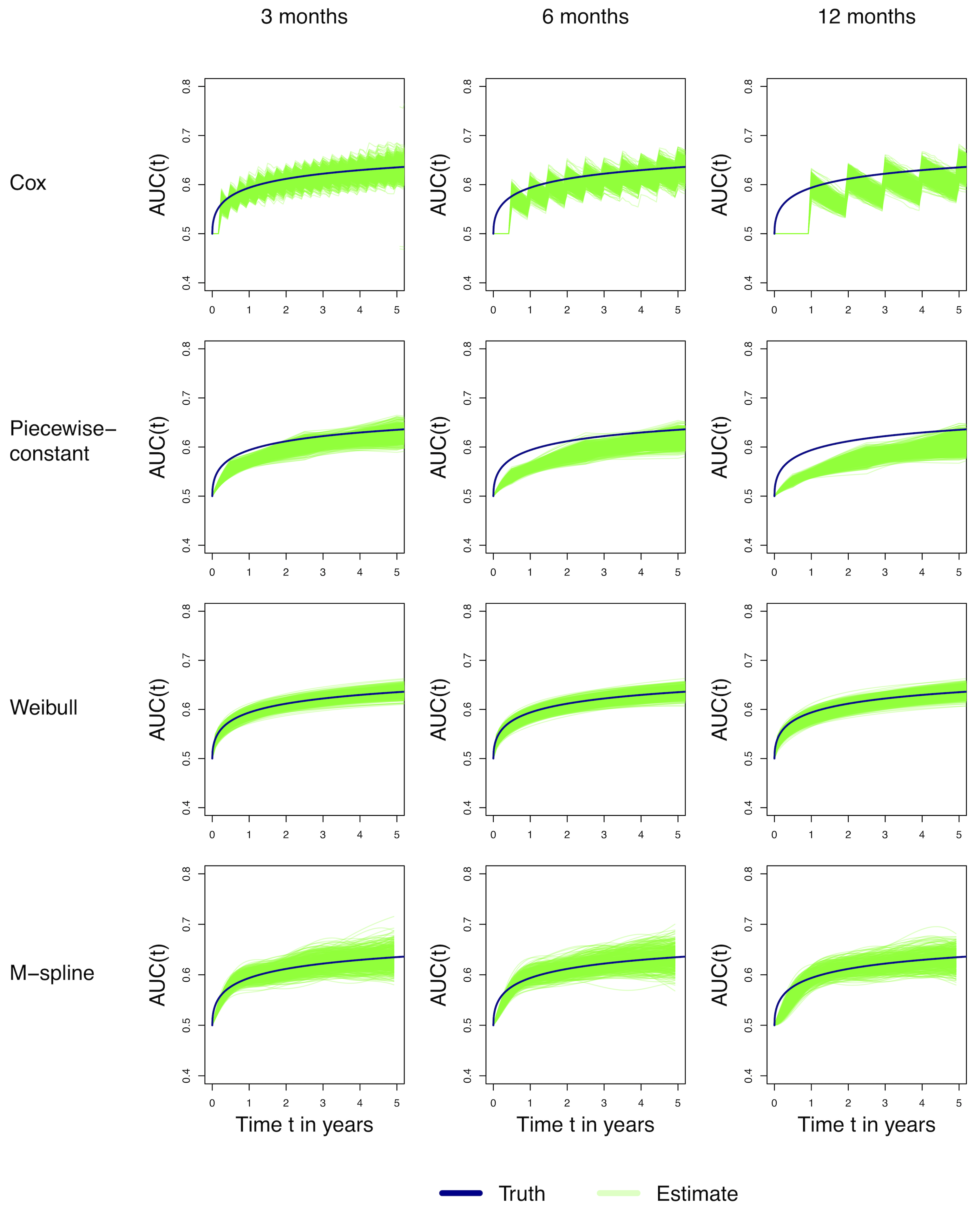}
	\caption{Estimated time-specific cumulative/dynamic AUC for scenario A (3 months; left panels), B (6 months; middle panels) and C (12 months; right panels) using different models (Cox, PW-const, Weibull, M-spline). The x-axis represents the prediction time $t$ in years. The prediction window is set to 5 years, so the y-axis represents $\widehat{\text{AUC}}^{\text{C/D}}(t,t+5)$. The blue line in each panel represents the true values over time.}
	\label{fig-auc-cd-abc}
\end{figure}

Figure \ref{fig-auc-cd-abc} shows the estimated cumulative/dynamic AUC 
over years with a prediction window of 5 years, i.e.,  $\widehat{\text{AUC}}^{\text{C/D}}(t,t+5)$, for scenarios A, B and C with follow-up visits every 3, 6, and 12 months, respectively (see Supplementary Figure C.2 for scenarios D-E-F). Each plot depicts the true AUC in blue and the AUC estimates of each simulated data set in green. 
For the M-spline model it is not possible to obtain transition probabilities for a time after the last observation time. This restricts the estimation of the cumulative/dynamic AUC to be estimated only until 5 years (i.e., the prediction window) prior to the last observation time.
The variation between AUC curves is much lower for the cumulative/dynamic AUC estimates in Figure \ref{fig-auc-cd-abc} compared to the incident/dynamic estimates in Figure \ref{fig-auc-id-abc}.
The Cox model displays jumps at the observation time points as in the incident/dynamic case, resulting in less bias at the times of follow-up visits compared to the times between follow-up visits.
The M-spline model's cumulative/dynamic curves underestimated the true AUC initially but recovered later on. The Weibull model outperformed the other models, but again one should keep in mind that data were generated from Weibull distributions.

\section{Application}\label{sec-application}

\subsection{Soft tissue sarcoma data}\label{sec-appl:sts}
\begin{figure}[h!]
	\centering
	\includegraphics[width=8cm]{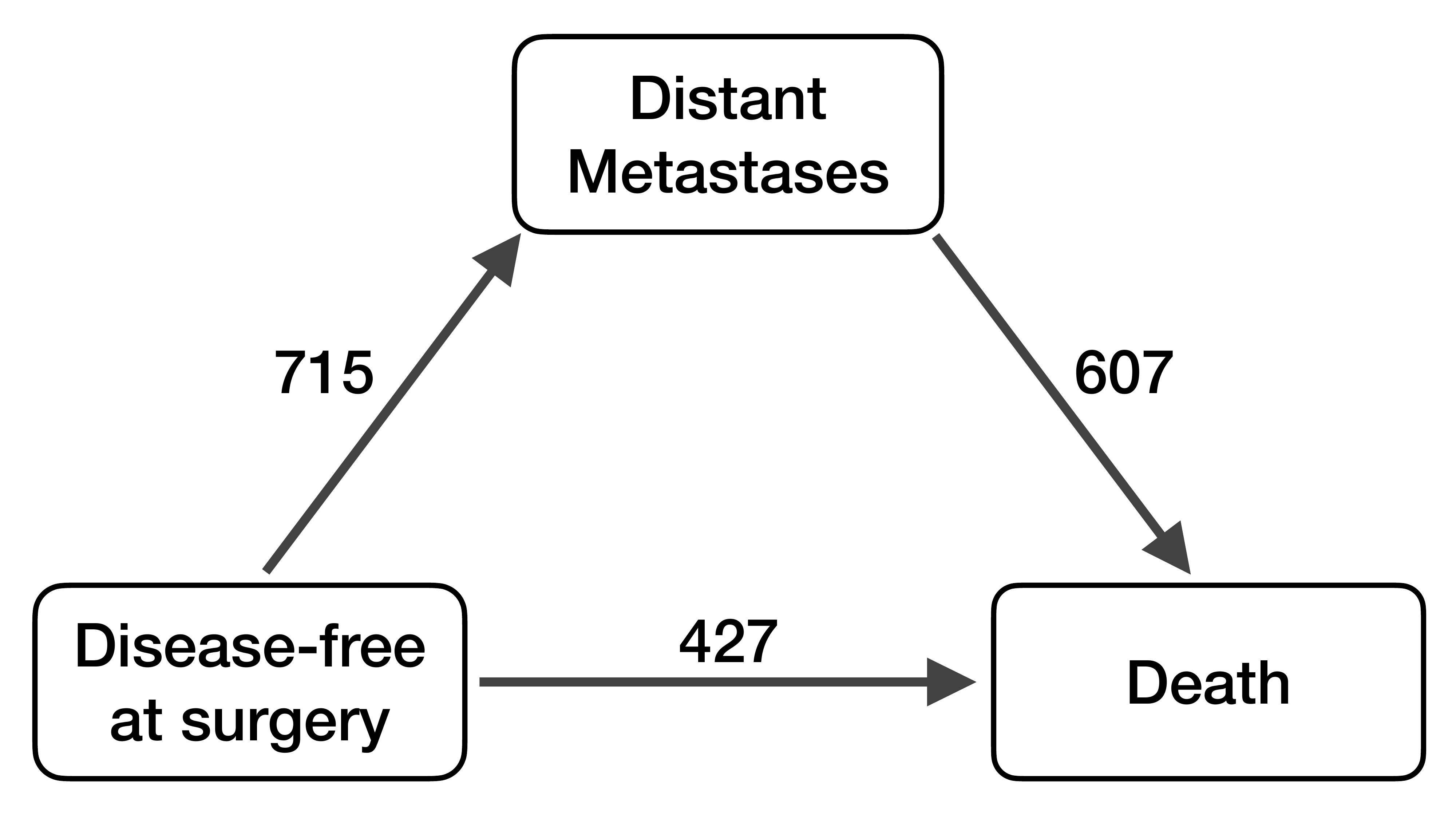}
	\caption{Soft tissue sarcoma illness-death model ($N$ = 2232). State 0 (disease-free): surgery with curative intent; state 1 (disease): development of distant metastases (DM); state 2: death. The numbers indicate the number of patients moving from one state to the other.}
	\label{fig-sts}
\end{figure}

The data analyzed in this section was used in \cite{dynamic} for the development of a dynamic prediction model for patients with high-grade soft tissue sarcoma. The data set contains follow-up information of 2232 patients treated surgically with curative intent. Median follow-up time was 6.42 years. After surgery disease progression can be described by several adverse events: a patient may develop a local recurrence and/or develop distant metastases (DM) and/or die. The analysis discussed in this section focuses on the effect of DM on death. In total 1034 patients died and 715 patients first developed DM (see Figure \ref{fig-sts}).

After surgery, a common scheme of follow-up visits for DM screening involves the patient being seen every 3 months during the first three years, then every 6 months until the fifth year, and once a year thereafter \citep{sts-guidelines}.
The data did not contain information about exact follow-up times and an approximation of disease screening times was applied to the data. For a patient who was diagnosed with DM during follow-up, the time of DM specified in the data was interpreted as the first positive screening for DM. Since the time of the last negative screening was unknown an approximate time of last screening was assumed: if DM was diagnosed within the first 3 years of follow-up, between 3 and 5 years, or after 5 years the previous screening was assumed to have taken place either 3, 6, or 12 months prior to DM diagnosis. 
A patient who was never diagnosed with DM was assumed to have been screened according to the common follow-up scheme described above. 

\subsection{Results}\label{sec-appl:results}
The four models presented in Section \ref{sec-IDM} were estimated for the soft tissue sarcoma data, and time-specific incident/dynamic and cumulative/dynamic AUCs introduced in Sections \ref{sec:ts-auc:id} and \ref{sec:ts-auc:cd}, respectively, were computed.

\begin{figure}[ht]
	\centering
	\includegraphics[width=0.9\linewidth]{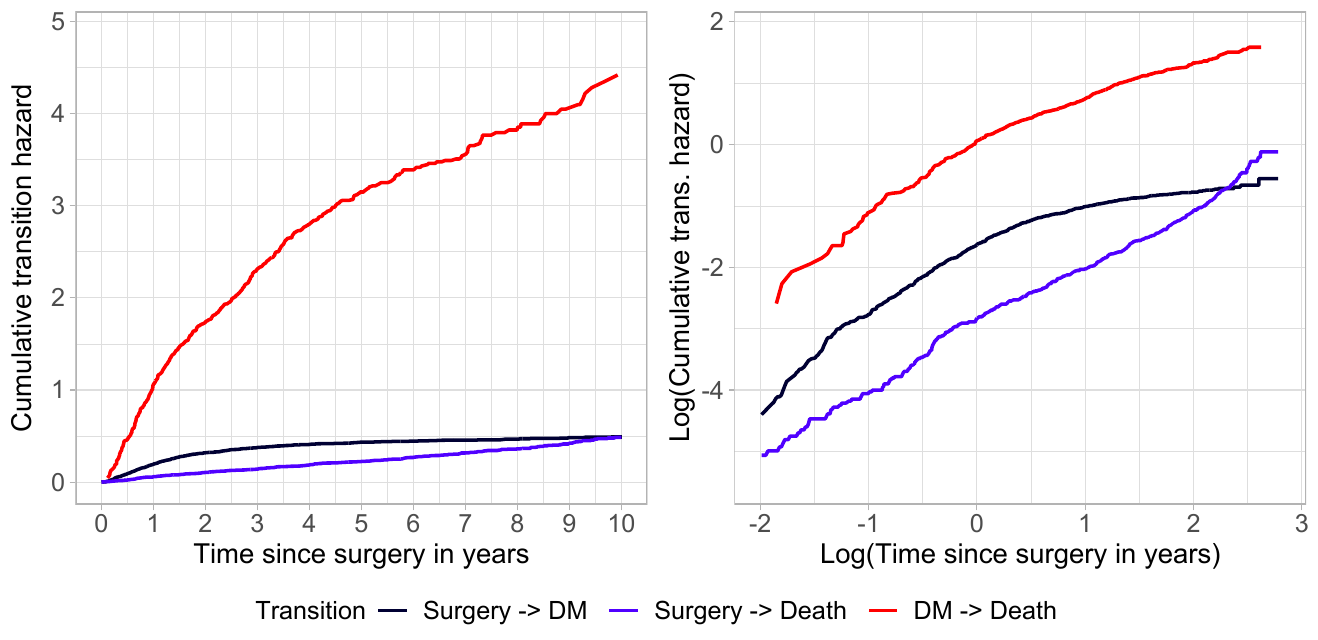}
	\caption{Left panel: Cumulative transition hazards. Right panel:  plot of logarithm of cumulative transition hazards versus logarithm of time (in years) to empirically check the fit of the Weibull distribution.}
	\label{fig-sts-haz}
\end{figure}

The HRs estimated by the Cox model and the piecewise-constant model were equal to 11.71 (95\% CI = [10.31; 13.29]) and 11.28 (95\% CI = [9.82; 12.96]) respectively. 
For the M-spline model, $m_{hl} = 5$ knots were used for each transition, and the smoothing parameters of the penalized likelihood were selected using the approximate leave-one-out cross-validation procedure implemented in the \texttt{SmoothHazard} package, by setting \texttt{idm(..., CV = 1)} \citep{sm-jss}. However, the M-spline model converged only after rescaling the time variable, whereas the parametric Weibull model converged without any rescaling.
Figure \ref{fig-sts-haz} displays on the left panel the non-parametric cumulative baseline hazards and on the right panel a graphical check of their fit to a Weibull distribution. For this figure the time of DM was assumed to be equal to the time that DM was detected during screening. If the data came from a Weibull distribution, the plot of the log-cumulative hazard against the log of time should be approximately linear (right panel), which is not the case in particular for the transition from surgery to DM (black line). The Weibull model therefore may not be appropriate for this data.

\begin{figure}[h!]
	\centering
	\includegraphics[width=\linewidth]{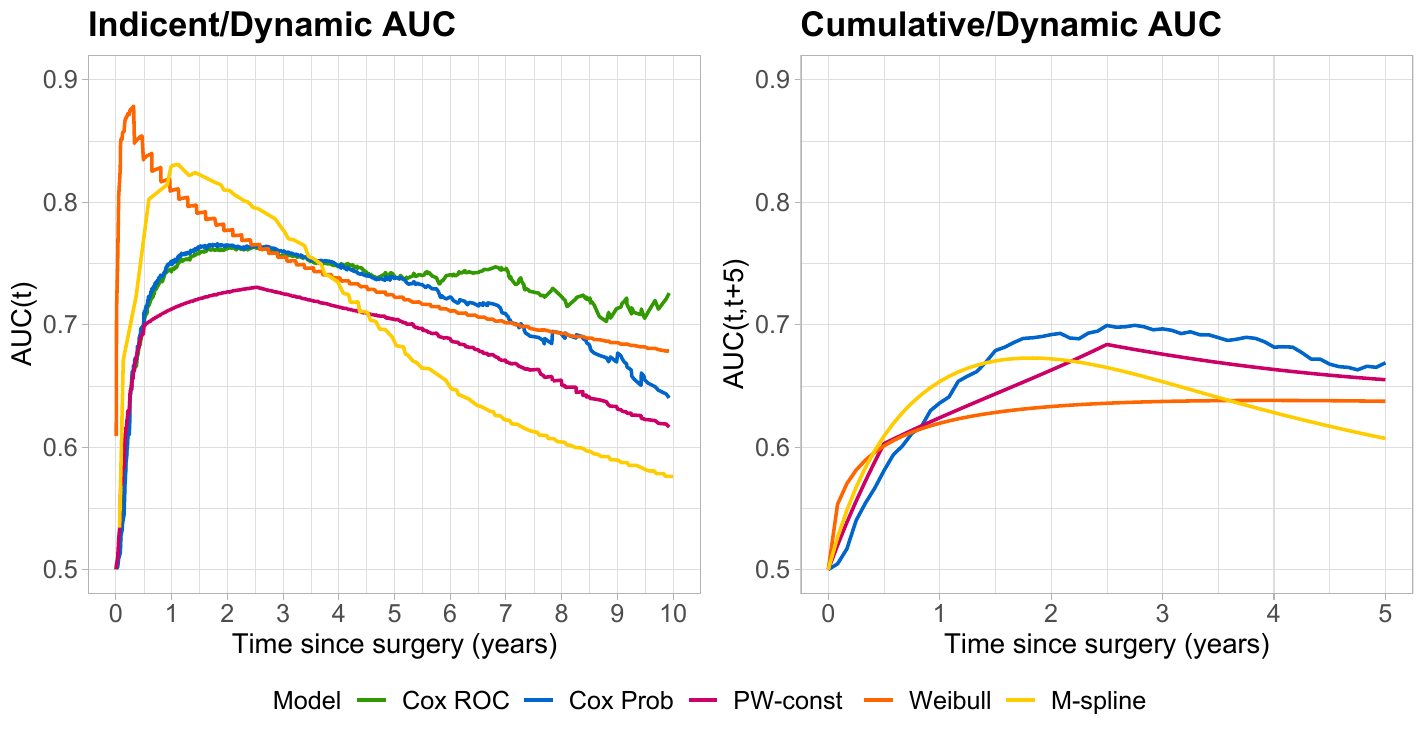}
	\caption{Time-specific AUC of the binary time-dependent marker $X(t)$ for the development of distant metastases in the soft tissue sarcoma cohort. Left panel: incident/dynamic AUC at year $t$. Right panel: cumulative/dynamic AUC for prediction time $t$ in years and prediction window of 5 years.\\
		Cox ROC (green): estimate based on Cox model \eqref{eq:cox} through \texttt{risksetAUC} function (only for incident/dynamic AUC). Cox Prob (blue): estimate  based on Cox model through \texttt{mstate} transition probabilities. PW-const (magenta): estimate based on piecewise-constant model \eqref{eq:pwc} through \texttt{msm}. Weibull (orange): estimate based on Weibull model \eqref{eq:weibull} through \texttt{SmoothHazard}. M-spline (yellow): estimate based on M-spline model \eqref{eq:mspline} with $m_{hl}=5$ knots per transition through \texttt{SmoothHazard}.}
	\label{fig-sts-auc}
\end{figure}

Figure \ref{fig-sts-auc} shows the AUCs over time for the different models (left panel: incident/dynamic; right panel: cumulative/dynamic), with time-specific values at years $t=1,2,3,4,5$ shown in Table \ref{tab-sts}. 
The incident/dynamic AUCs of the Weibull (orange) and M-spline (yellow) models are initially much higher than those of the other models and decline over time, suggesting better prognostic ability in the short term than in the long term.
The incident/dynamic AUC of the piecewise-constant model (magenta) is the lowest among all methods up to 5 years, after which the M-spline performs worse.
The cumulative/dynamic AUC of the Cox model (blue) is the largest from $t=1.5$ years after surgery onwards, while the Weibull model performs worst, except at the beginning of follow-up. 
The M-spline shows good performance in the short term but deteriorates over time.

\begin{table}[h!]
	\small\centering\caption{Estimated discrimination performance of the binary time-dependent marker $X(t)$ for the development of distant metastases (DM) in the soft tissue sarcoma cohort. $\text{AUC}^{\text{I/D}}(t)$: incident/dynamic AUC at year $t$. $\text{AUC}^{\text{C/D}}(t,t+5)$: cumulative/dynamic AUC for prediction time $t$ in years and prediction window of 5 years.}\label{tab-sts}
	\begin{tabular}{ccccccc}
		\toprule
		Time-specific AUC & Year $t$ & Cox ROC & Cox Prob & PW-const & Weibull & M-spline \\ 
		\midrule
		$\widehat{\text{AUC}}^{\text{I/D}}(t)$ & 1 & 0.74 & 0.75 & 0.71 & 0.81 & 0.83  \\ 
		& 2 & 0.76 & 0.76 & 0.73 & 0.78 & 0.81 \\ 
		& 3 & 0.76 & 0.76 & 0.73 & 0.76 & 0.78  \\ 
		& 4 & 0.75 & 0.75 & 0.71 & 0.74 & 0.73 \\
		& 5 & 0.74 & 0.74 & 0.70 & 0.72 & 0.69  \\
		\midrule
		$\widehat{\text{AUC}}^{\text{C/D}}(t,t+5)$ & 1 &      & 0.64 & 0.62 & 0.62 & 0.65  \\
		& 2 &      & 0.69 & 0.66 & 0.63 & 0.67 \\
		& 3 &      & 0.70 & 0.68 & 0.64 & 0.65  \\
		& 4 &      & 0.68 & 0.66 & 0.64 & 0.63  \\
		& 5 &      & 0.67 & 0.66 & 0.64 & 0.61  \\
		\bottomrule
		\multicolumn{7}{p{13cm}}{\footnotesize \textbf{Note:}  \textit{Cox ROC}, estimate based on Cox model \eqref{eq:cox} through \texttt{risksetAUC} function (only for incident/dynamic AUC); \textit{Cox Prob}, estimate  based on Cox model through \texttt{mstate} transition probabilities; \textit{PW-const}, estimate based on piecewise-constant model \eqref{eq:pwc} through \texttt{msm}; \textit{Weibull}, estimate based on Weibull model \eqref{eq:weibull} through \texttt{SmoothHazard}. \textit{M-spline}, estimate based on M-spline model \eqref{eq:mspline} with $m_{hl}=5$ knots per transition through \texttt{SmoothHazard}.}
	\end{tabular}
\end{table}

\section{Discussion}\label{sec-discussion}

The illness-death model is frequently applied to clinical data to describe disease progression. A patient enters the model disease free, may then experience disease and die. 
In clinical practice, however, the time of disease onset cannot often be observed exactly. The information is interval-censored or not observed due to death or censoring. This can lead to bias in the estimation of disease incidence and regression coefficients \citep{Cook2020,Commenges2002,joly2002, leffondre}.

This article studied the discrimination performance of a binary time-dependent disease marker in the context of the illness-death model for interval-censored data. A simulation study with several data scenarios was conducted to study four different models: the Cox model with disease as time-dependent marker, the piecewise-constant model implemented in the \texttt{msm} package, the Weibull model and the M-spline model implemented in the \texttt{SmoothHazard} package. 
Both incident/dynamic and cumulative/dynamic AUC estimates were derived from their transition probabilities and studied.
The methods were also applied to a data set of patients surgically treated for soft tissue sarcoma who were scanned for distant metastases at scheduled follow-up visits. 
In both cases, covariates were not included in the models, but they could be incorporated on a transition-specific basis, allowing each transition $h \rightarrow l$ to have distinct covariate effects, modeled multiplicatively on the hazard as $\exp(\boldsymbol{\beta}^T_{hl}\boldsymbol{Z}_{hl})$.

The simulation study showed that the HRs from the piecewise-constant model were less biased than those of the Cox model. 
The number of patients per data set (400 vs 1000 vs 2000) did not have a large effect on the estimates of the HR, AUC estimates in incident/dynamic and cumulative/dynamic definition, except for the M-spline model which converged more reliably with large data sets (see Supplementary Material C). 
The type of censoring scheme did mostly influence AUC estimates based on the Cox model and M-spline model, since hazard estimates are non-parametric and therefore more sensitive at later time points at which fewer events are observed.
The Weibull model demonstrated the best performance; however, it had an inherent advantage, as the simulated data followed a Weibull distribution. In practice a Weibull distribution may not be a good fit to the data, as shown in the soft tissue sarcoma application.
The M-spline model showed a good performance when estimating the incident/dynamic and cumulative/dynamic AUC however was not always able to converge and provide AUC estimates. 
AUC estimates based on the piecewise-constant model in the incident/dynamic definition had less bias than those based on the Cox model for scenarios with large spacing between follow-up visits and in the cumulative/dynamic definition they had the largest bias of all methods.
The spacing of follow-up visits at which the disease state was observed did have a large effect on estimates of the incident/dynamic AUC, particularly for the Cox model.
The cumulative/dynamic AUC depends solely on transition probabilities, whereas the incident/dynamic AUC also depends on the underlying transition hazards. Consequently, discrimination measured by the cumulative/dynamic AUC is largely insensitive to differences in baseline hazard specification. This explains why the considered methods do not generally outperform the Cox-based approach for cumulative/dynamic AUC, particularly at follow-up visit times when the Cox model exhibits lower bias.

In practice, depending on how reasonable the assumption of Weibull distributed transition hazards is, one could choose to estimate the AUCs based on the Weibull or M-spline model.
However, it is crucial to consider that convergence of M-spline models can be affected by smoothing parameter selection or numerical instability. Automatic smoothing parameter selection via cross-validation, while generally effective, can be computationally intensive, and manual tuning is often challenging in complex datasets and not always feasible in simulation studies. In our application, rescaling the time variable enabled convergence under automatic smoothing parameter selection, highlighting the importance of careful numerical considerations in spline-based modeling. Specifically, extreme values or a wide range of time points can lead to poorly scaled basis functions and penalty terms, complicating optimization and potentially preventing convergence.

The choice between incident/dynamic and cumulative/dynamic discrimination metrics depends on the specific perspective of interest, yet both are scientifically relevant. Incident/dynamic AUC is most appropriate when the focus is on the marker's ability to identify, at a given time, which individuals who are still at risk are most likely to experience death shortly thereafter. This can be particularly informative for time-specific decision-making, such as screening or monitoring. In contrast, cumulative/dynamic AUC evaluates the marker’s ability to discriminate which individuals who are still at risk at a given time are most likely to experience death before a given time horizon, thereby capturing its long-term prognostic capacity. Since clinical and research objectives may require either short-term or cumulative assessments, considering both approaches provides a more comprehensive evaluation of marker performance.

A limitation of multi-state models, such as the illness-death model, is that small sample sizes (e.g., 100 or 200 subjects) often lead to unreliable estimates, even when event times are exactly observed, due to the complexity of modelling multiple transitions and the potential sparsity of events. This issue is further amplified in the presence of interval censoring. Consequently, caution is warranted when applying the proposed methods in such settings.
Moreover, it was implicitly assumed that the visiting process did not depend on the state (diseased, non-diseased). In clinical practice however, this may not always be a reasonable assumption. Patients may visit the clinic earlier if they have complaints that are related to being in the diseased state. The effect of an informative visiting process could be a subject of future research.

The performed simulations examined the effect of an interval-censored binary disease marker. Future research could explore the discriminatory performance of an interval-censored, time-dependent covariate with more than two possible values, i.e., a subject that can transition between multiple disease states. 
This would require estimation methods for general interval-censored multi-state models, which have been developed only in recent years \citep{Machado2018,Gu2024,Gomon2025}.
Another topic of future research could be to investigate a different definition of the incident/dynamic AUC. At time $t$ it was defined as the probability that the individual that dies at time $t$ was diseased at time $t$; however one could also consider an incident/dynamic AUC in which the disease state is evaluated at an earlier time point $s$, where $s < t$ \citep{zheng2004}.

This study highlights the importance of considering the interval-censored nature of disease data in both parameter estimation and the evaluation of discrimination performance for disease development. This consideration is crucial as prediction models are nowadays increasingly important in clinical practice to provide personalized patient care.


\vspace{0.5cm}
\subsubsection*{Supplementary Material} Supplementary Material is available online in \textit{Statistical Methods in Medical Research} along with the published article (DOI: \href{https://journals.sagepub.com/doi/10.1177/09622802251412855}{10.1177/09622802251412855}).

\newpage
\small

\noindent\textbf{Funding.}
A.J. R.-B. was supported by the KWF Kankerbestrijding grant UL2015-8028. M.S. was supported by the KWF Kankerbestrijding grant 2023-3 DEV/15461. The funding sources had no involvement in study design, in the collection, analysis and interpretation of the data, in writing of the report, and in the decision to submit the article for publication.

\vspace{0.5cm}
\noindent\textbf{Acknowledgments.}
The simulation study was performed using the compute resources from the Academic Leiden
Interdisciplinary Cluster Environment (ALICE) provided by Leiden University. 
Prof.dr. Michiel van de Sande and the PERsonalized SARcoma Care (PERSARC) study group are gratefully acknowledged for making the soft tissue sarcoma data set available. \\
\textit{PERSARC study group:}
Lee M. Jeys, Minna K. Laitinen, Rob Pollock, Will Aston, Jos A. van der Hage, Sander Dijkstra, Peter C. Ferguson, Anthony M. Griffin, Julie J. Willeumier, Jay S. Wunder, Emelie Styring, Florian Posch, Olga Zaikova, Katja Maretty-Kongstad, Johnny Keller, Andreas Leithner, Maria A. Smolle, Rick L. Haas.

\vspace{0.5cm}
\noindent\textbf{Data availability statement.}
The soft-tissue sarcoma data are not publicly available due to privacy restrictions. 
The R code used to generate the simulated data sets as described in Section~\ref{sec-simu-data} and perform the simulation study is provided at \url{{http://github.com/mspreafico/auc-idmIC}}.

\end{document}